\documentclass[useAMS,usenatbib,usegraphicx]{mn2e}

\def\nodata{...}
\def\p1{\phantom{1}}
\def\p0{\phantom{0}}

\def\simless{\mathbin{\lower 3pt\hbox
     {$\rlap{\raise 5pt\hbox{$\char'074$}}\mathchar"7218$}}}   
\def\simmore{\mathbin{\lower 3pt\hbox
     {$\rlap{\raise 5pt\hbox{$\char'076$}}\mathchar"7218$}}}   
\def\Msun{{\rm M}_\odot}                                       
\def\hide#1{}

\title[Lags of the HFQPOs in four black-hole systems]{The phase lags of high-frequency quasi-periodic oscillations in four black-hole candidates}
\author[M. M\'endez, D. Altamirano, T. Belloni \& A. Sanna]
{Mariano M\'endez$^{1}$\thanks{mariano@astro.rug.nl}, 
Diego Altamirano$^{2}$, 
Tomaso Belloni$^{3}$, 
Andrea Sanna$^{4}$\\ 
$^{1}$Kapteyn Astronomical Institute, University of Groningen, P.O. Box 800, 9700 AV Groningen, The Netherlands\\
$^{2}$Astronomical Institute Anton Pannekoek, University of Amsterdam, Science Park 904, 1098 XH Amsterdam, The Netherlands\\
$^{3}$INAF - Osservatorio Astronomico di Brera, Via E. Bianchi 46, I-23807, Merate, Italy\\
$^{4}$Dipartimento di Fisica, Universit\`a degli Studi di Cagliari, SP Monserrato-Sestu km 0.7, 09042 Monserrato, Italy}

\begin{document}

\date{Accepted 2013 July 31.  Received 2013 July 10; in original form 2013 June 6}

\pagerange{\pageref{firstpage}--\pageref{lastpage}} \pubyear{2013}

\maketitle

\label{firstpage}

\begin{abstract}
We measured the phase-lag spectrum of the high-frequency quasi-periodic oscillations (QPO) in the black hole systems (at QPO frequencies) GRS 1915+105 (35 Hz and 67 Hz), GRO J1655--40 (300 Hz and 450 Hz), XTE J1550--564 (180 Hz and 280 Hz), and IGR J17091--3624 (67 Hz). The lag spectra of the 67-Hz QPO in, respectively, GRS 1915+105 and IGR J17091--3624, and the 450-Hz QPO in GRO J1655--40 are hard (hard photons lag the soft ones) and consistent with each other, with the hard lags increasing with energy. On the contrary, the lags of the 35-Hz QPO in GRS 1915+105 are soft, with the lags becoming softer as the energy increases; the lag spectrum of the 35-Hz QPO is inconsistent with that of the 67-Hz QPO. The lags of the 300-Hz QPO in GRO J1655--40, and the 180-Hz and the 280-Hz QPO in XTE J1550--564 are independent of energy, consistent with each other and with being zero or slightly positive (hard lags). For GRO J1655--40 the lag spectrum of the 300-Hz QPO differs significantly from that of the 450-Hz QPOs. The similarity of the lag spectra of the 180-Hz and 280-Hz QPO in XTE J1550--564 suggests that these two are the same QPO seen at a different frequency in different observations. If this is correct, the lags could provide an alternative way to identify the high-frequency QPO features in galactic black-hole systems. The lag spectrum of the 67-Hz QPO in GRS 1915+105 (hard lags) is significantly different from that of the $2.7 \times 10^{-4}$ Hz QPO in the narrow-line Seyfert 1 galaxy RE J1034+396 (soft lags), which disproves the suggestion that the two QPOs are the same physical phenomenon with their frequencies scaled only by the black-hole mass. The lag spectrum of the QPO in RE J1034+396 is similar to that of the 35-Hz QPO in GRS 1915+105, although identifying these two QPOs as being the same physical feature remains problematic. We compare our results with those for the lags of the kilohertz QPOs in neutron-star systems and the broadband noise component in active galactic nuclei, and discuss possible scenarios for producing the lags in these systems.
\end{abstract}


\section{Introduction}
\label{intro}

In addition to low frequency variability \citep[see e.g.,][and references therein]{vanderklis01,remillard-mcclintock,belloni2010,motta11}, the X-ray light curves of galactic black-hole systems show quasi-periodic variability at frequencies of $\sim30-450$ Hz \citep[][see \citealt{belloni-sanna-mendez} and references therein]{Morgan:1997hn, remillard99,belloni-mendez-sanchez,homan,miller2001,strohmayer,strohmayer02,remillard02a,altamirano-belloni}. This variability evinces as relatively narrow features in the Fourier power density spectra of these sources, the so-called high-frequency quasi-periodic oscillations (HFQPOs). The mean frequency of some of these HFQPOs dovetails with the Keplerian orbital frequency at a few gravitational radii of an object of $10-20~\Msun$, and therefore these HFQPOs have attracted much attention \citep[e.g.,][]{nowak02,cui02,fragile}. 

From more than 7000 observations of black-hole systems with the {\em Rossi X-ray Timing Explorer (RXTE)}, there are only about a dozen reports in the literature of HFQPOs \citep*[see references in][]{belloni-sanna-mendez}. In several cases the QPOs reported were marginally significant; a systematic analysis carried out by \cite*{belloni-sanna-mendez} using individual observations for a total of $\sim12$ Ms of {\em RXTE} data from 22 galactic black hole systems, and excluding the peculiar black-hole systems GRS 1915+105 and IGR J17091--3624, yielded only 11 significant detections of HFQPOs from two sources, XTE J1550--564 and GRO J1655--40. \citep*[A handful of the original detections of HFQPOs were based on combinations of observations within specific states of the source, or specific energy selections, and hence those QPOs are not detected in the procedure used by][]{belloni-sanna-mendez}. Remarkably, the fraction of observations with HFQPOs in black-hole systems is dramatically lower than in neutron-star systems, where QPOs at $400-1200$ Hz are usually detected in $\sim40$\% of the observations \citep[see, e.g.,][for the neutron-star system 4U 1636--53]{sanna02}.

GRS 1915+105 is an exceptional case. Besides displaying a wide sweep of variability \citep{belloni00}, it has shown several HFQPOs, at frequencies of 27 Hz \citep*{belloni-mendez-sanchez}, 35 Hz \citep{belloni-altamirano}, 41 Hz \citep{strohmayer02} and 67 Hz \citep*{Morgan:1997hn,belloni-altamirano2}.

Besides XTE J1550--564, GRO J1655--40, and GRS 1915+105, the transient source IGR J17091--3624 has also recently shown HFQPOs. In many respects this source appears to be a scaled-down version of GRS 1915+105 \citep{altamirano}, including the presence of a HFQPO at 67 Hz \citep{altamirano-belloni}, the same exact frequency (within errors) as the highest (and strongest) HFQPO seen in GRS 1915+105.

One interesting aspect of the variability of a signal in general, and of the HFQPOs in particular, is the possible phase lag and coherence between measurements in two different energy bands \citep[e.g.,][]{nowak01}. The phase lag is a Fourier-frequency-dependent measure of the phase delay between two concurrent and correlated signals, in this case light curves of the same source, in two different energy bands. The coherence measures the frequency-dependent degree of linear correlation between the two signals. Perfect (unit) intrinsic (noise subtracted) coherence entails that the two signals are related through a linear transfer function \citep[see][for a detailed illustration of the use of the coherence function in X-ray astronomy; see also \citealt{bendat01} for a comprehensive explication of transfer functions in linear systems]{vaughan01,nowak01}.

%
%
%
\begin{table*}
\caption{Energy bands used to calculate the phase lags of the HFQPOs in GRS 1915+105, GRO J1655--40, XTE J1550--564, and IGR J17091--3624. We indicate with an asterisk the band that we used as reference to calculate the phase lags and coherence.}
\centering
\begin{tabular}{cccc}
\hline
GRS 1915+105 & GRO J1655--40 & XTE J1550--564 & IGR J17091--3624 \\
\hline
$E$ [keV] (channels) & $E$  [keV] (channels) & $E$ [keV] (channels) & $E$ [keV] (channels) \\
\hline
$\phantom{0^*}3.2 - \phantom{0}5.7 ~ (\phantom{0}8-13)^* $ & $\phantom{13.0^*} < 13.0 ~ (\phantom{0}0-35)^*$ & $\phantom{13.0^*} <   \phantom{0}6.5 ~ (\phantom{0}0-17)^*$ & $\phantom{13.0} < \phantom{0}3.2 ~ (\phantom{0}0-\phantom{11}7)$ \\
$\phantom{0}5.7 -                   14.8 ~                   (14-35) $ & $13.0 - 15.2  ~                  (36-41)$ & $ \phantom{0}6.5 - 13.0 ~ (18-35)$ & $\phantom{0^*}3.2 - \phantom{0}5.7 ~ (\phantom{0}8-\phantom{1}13)^*$ \\
$                  14.8 -                   17.3 ~                    (36-41) $ & $15.2 - 18.1 ~                  (42-49)$ & $13.0 - 15.2 ~ (35-41)$ &$\phantom{0}5.7 -                     14.8 ~                     (14-\phantom{1}35) $\\
$                  17.3 -                   20.6 ~                    (42-49) $ & $18.1 - 26.3 ~                  (50-71)$ & $15.2 - 18.1 ~ (42-49)$ &$                    14.8 -                   43.0 ~                    (36-100) $\\
$                  20.6 -                   30.0 ~                    (50-71) $ & $\nodata          ~                              $ & $18.1 - 26.2 ~ (50-71)$ & $\nodata $\\
\hline
\end{tabular}
\label{tab1}
\end{table*}

The phase-lag spectrum of the broadband noise component in the power density spectrum of black-hole systems \citep[e.g.][]{miyamoto01, nowak01} has been interpreted as due to soft disc photons that are Compton up-scattered in a corona of hot electrons that surrounds the system \citep[e.g.,][]{kazanas01,hua01}. The extension of the measurements of phase lags to the HFQPOs in black-hole system provides a powerful tool to study the physics of the accretion flow in the vicinity of the black hole. The only measurement of the phase-lag spectrum of a HFQPOs is that of \cite{cui}, on the 67-Hz QPO in GRS 1915+105. \cite{cui} found that the phase lags are hard (the hard photons lag the soft ones) and increase with energy. The corresponding time lags (equal to the phase lags divided by $2\pi$ the frequency of the QPO) reached up to $\sim5$ ms. Using {\em RXTE} data, in this paper we measured the phase-lag spectrum and the Fourier coherence of the HFQPOs of the four black-hole systems GRS 1915+105, XTE J1550--564, GRO J1655--40, and IGR J17091--3624. We describe the observations and data analysis in \S\ref{obs}; in \S\ref{res} we present our results, and we discuss them in \S\ref{dis}.

\section{Observations and Data Analysis}
\label{obs}

We analysed archival data of the four black-hole X-ray binaries GRS 1915+105, GRO J1655--40, XTE J1550--564, and IGR J17091--3624, collected with the Proportional Counter Array \citep[PCA;][]{jahoda01} on board {\em RXTE} \citep[][]{bradt01}. We always used the highest time-resolution data available in each observation to calculate Fourier transforms of light curves in different energy bands (see below). Due to the ageing of the PCA instrument, the channel-to-energy conversion changed over time. Those changes took place over relatively long time-scales (of the order of years; see for instance \citealt{straaten01}), except for a few occasions in which the mission control changed the gain of the PCA instrument manually. These manual gain changes define the so-called PCA gain epochs\footnote{The definition of the PCA gain epochs and the channel-to-energy conversion table for each epoch are available at http://heasarc.gsfc.nasa.gov/docs/xte/r-c\_table.html}. 

For GRS 1915+105 we took the observations in \citet{belloni-altamirano}, in which the QPO at $\sim67$ Hz was very significant. In these observations \citet{belloni-altamirano} also detected a second HFQPO at 35 Hz. We will here call these the 67-Hz and 35-Hz QPO, respectively. These observations were done on MJD 52933.627, 52933.696 and 52933.764, within the same gain epoch of the PCA (gain epoch 5). This allowed us to use the same channel scheme to combine the data of the three observations for the same energy intervals (see below). 

For GRO J1655--40 and XTE J1550--564 we used the observations in which \citet*{belloni-sanna-mendez} detected significant HFQPOs. From their Table 2, for GRO J1655--40 we selected the observations on MJD 50296.311, 50324.280, 50335.913, and 50394.884, with HFQPOs between $\sim270$ Hz and $\sim450$ Hz, and all taken within PCA gain epoch 3. 

For XTE J1550--564 we used the observations from MJD 51076.000, 51108.076, 51241.802, 51242.507 and 51255.158, with HFQPOs between $\sim180$ Hz and $\sim280$ Hz, since these observations were all taken within the same PCA gain epoch (gain epoch 3), and we dropped the last two observation in the list of \citet*{belloni-sanna-mendez} which were taken in PCA gain epoch 4. 

For IGR J17091--3624, following \citet{altamirano-belloni}, we used 44 observations with a HFQPO at $\sim67$ Hz, taken in the period MJD 55830 to 55880, all of them from PCA gain epoch 5. These observations correspond to the low-variability period marked with a grey area in Figure 1 of \citet{altamirano-belloni}. We will call this the 67-Hz QPO in IGR J17091--3624. We give the channel and energy boundaries of the bands used for each source in Table \ref{tab1}.

For each observation we computed the complex Fourier transform for all photons available, and separately for those in each energy band in Table \ref{tab1}, choosing a Nyquist frequency of 1024 Hz over segments of 16 s. From the Fourier transforms we calculated power density spectra over the full energy band. We normalised all power density spectra following \citet{leahy}, and finally we produced an average power density spectrum for each observation and each source separately.

Since the HFQPO at $\sim450$ Hz in GRO J1655--40 is only detected above 13 keV \citep*{strohmayer, belloni-sanna-mendez}, for this source we also created a power density spectrum of the observation with this QPO using only photons in the $13-26.3$ keV energy range. For IGR J17091--3624, following \cite{altamirano-belloni}, we produced power density spectra in the $2-25$ keV band. 

We fitted each power density spectrum from 20 Hz below up to 100 Hz above the QPO central frequency reported in the literature, using a model that consisted of a constant term to represent the Poisson noise, one Lorentzian for the HFQPO (in the case of GRS 1915+105 we fitted two Lorentzians to the two HFQPOs that were present simultaneously in the data) and a zero-centred Lorentzian to fit the contribution of the low-frequency broadband noise component to the frequency range of the HFQPO. 

When we had more than one observation of a source with a HFQPO, we averaged the power density spectra of those observations. However, before doing this last step, (i) when there was more than one HFQPO in an observation, or (ii) when we detected a single HFQPO with a frequency that was not the same in different observations, we first had to identify the HFQPOs and, if necessary, correct for their frequency variations (see below), before averaging the power density spectra. For each source and each QPO, we did this as follows: 

(i) The frequency of the 67-Hz QPO in GRS 1915+105 was consistent with being the same in all the observations. The same is true for the 67-Hz QPO in IGR J17091--3624. We therefore assumed that each of these two HFQPOs was the same physical feature in, respectively, all three observations of GRS 1915+105 and all 44 observations of IGR J17091--3624. 

In GRS 1915+105, the other HFQPO was in all cases consistent with having a constant frequency of 35 Hz, therefore we also assumed that the 35-Hz HFQPO was always the same physical QPO in the three observations. 

In three of the four observations of GRO J1655--40 with a statistically significant HFQPO \citep*[see Table 2 in][]{belloni-sanna-mendez}, the frequency of the HFQPO was at $\sim270$ Hz, $\sim290$ Hz, and $\sim310$ Hz, respectively, while in the remaining observation the HFQPO was at $\sim450$ Hz. Following the identification of \cite{strohmayer}, we took the HFQPOs at $\sim270-310$ Hz and the one at $\sim450$ Hz as two different QPOs. We call these the 300-Hz and the 450-Hz QPO, respectively. 

In two out of the five observations of XTE J1550--564, the HFQPO was at $\sim180$ Hz, whereas in the other three observations the HFQPO was at $\sim280$ Hz \cite*[Table 2 in][]{belloni-sanna-mendez}. We call these the 180-Hz and the 280-Hz QPO, respectively; we assumed that the 180-Hz and 280-Hz HFQPOs were two different QPOs, and analysed them separately. 

(ii) In the case of GRS 1915+105, since the 35-Hz and 67-Hz QPOs had the same frequency in the three observations, we simply combined the data. 

Similarly, in the case of IGR J17091--3624 the 67-Hz QPO was always at around the same frequency, and therefore we combined the data directly. 

Since we had only one observation of GRO J1655--40 with the 450-Hz QPO, we only used the data of that observation in our analysis of this HFQPO. 

For the other three observations of GRO J1655--40, and for the five observations of XTE J1550--564, we applied the shift-and-add technique described in \citet{mendez01}: We shifted the frequency scale of the power density spectra of the individual observations such that the frequency of the HFQPO was the same and we averaged the power density spectra. 

%
%
%
\begin{table}
\caption{Average frequency, $\nu_{\rm HFQPO}$, and $FWHM$ of the HFQPOs in GRS 1915+105, GRO J1655--40, XTE J1550--564, and IGR J17091--3624. Except for the $\sim450$-Hz QPO in GRO J1655--40 and the 67-Hz QPO in IGR J17091--3624, we measured the frequency and $FWHM$ of the HFQPOs in power density spectra calculated over the full energy band. To measure the frequency and $FWHM$ of the $\sim450$-Hz QPO in GRO J1655--40 we used a power density spectrum calculated between 13 keV and 26.3 keV. For IGR J17091--3624 we calculated the power density spectrum between 2 keV and 25 keV.}
\centering
\begin{tabular}{ccc}
\hline
& $\nu_{\rm HFQPO}$ [Hz] & $FWHM$ [Hz]\\
\hline
GRS 1915+105                   & $\p0 35.1 \pm 0.4 $ & $\p0\p0\p0\p0 4.2 \pm 1.4$\\
                                            & $\p0 67.9 \pm 0.1 $ & $\p0\p0\p0\p0 2.3 \pm 0.2$ \\
GRO J1655--40                   & $289 \pm 4$           & $\p0\p0\p0104 \pm 13$ \\
                                            & $446 \pm 2$           & $\p0\p0\p0\p051 \pm 30$ \\
XTE J1550--564                  & $182 \pm 3$           & $\p0\p0\p0\p066 \pm 12$ \\
                                            & $282 \pm 3$           & $\p0\p0\p0\p032 \pm 10$ \\
IGR J17091--3624              & $\p0 66.5 \pm 0.5 $ & $\p0\p0\p0\p0 8.5 \pm 1.6$ \\
                                            
\hline
\hline
\end{tabular}
\label{tab2}
\end{table}

\begin{figure*}
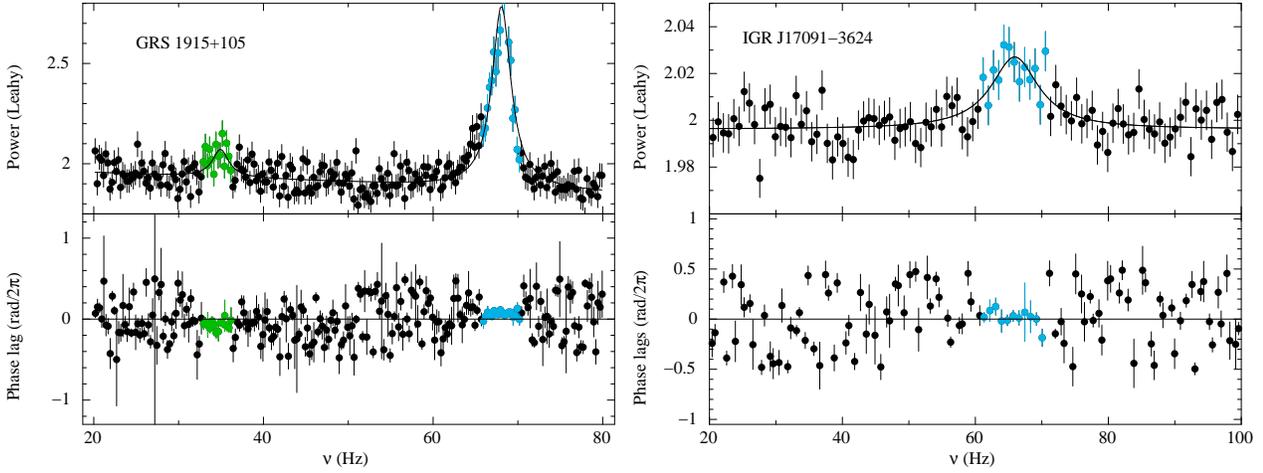

\centering
\includegraphics[width=0.35\textwidth,angle=-90]{figure1a.ps}
\includegraphics[width=0.35\textwidth,angle=-90]{figure1b.ps}
\caption{The 35-Hz and 67-Hz QPO in GRS 1915+105 (left panels) and the 67-Hz QPO in IGR J17091--3624 (right panels). The upper panels show the power density spectra of the full PCA bandpass light curve for GRS 1915+105 and for photons in the $2-25$ keV band for IGR J17091--3624. The lower panels show the phase-lag Fourier spectra (phase lags in radians divided by $2\pi$ vs. Fourier frequency) for the photons in the $5.7-14.8$ keV band relative to those in the $3.2-5.7$ keV band. The green and the light-blue points in the left panels show, respectively, the frequency range over which we calculated the phase lags of the 35-Hz and 67-Hz HFQPO in GRS 1915+105. The light-blue points in the right panels show the frequency range over which we calculated the phase lags of the 67-Hz HFQPO in IGR J17091--3624 (see text for details). The solid line in the upper panels shows the best-fitting model to this part of the power-density spectrum. The horizontal line in the lower panels shows the zero-lag level.}
\label{fig1}
\end{figure*}

\begin{figure*}
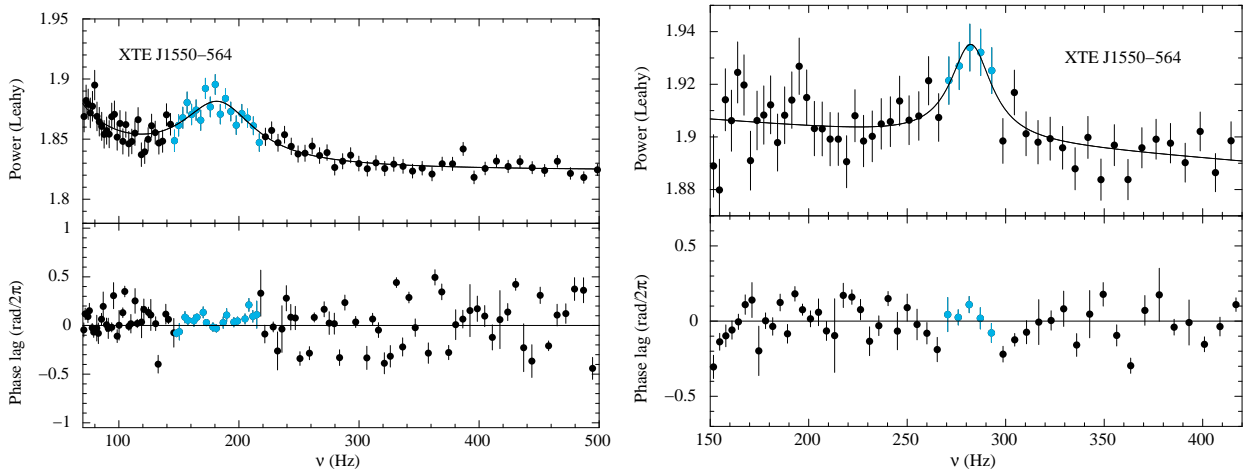

\centering
\includegraphics[width=0.35\textwidth,angle=-90]{figure2a.ps}
\includegraphics[width=0.35\textwidth,angle=-90]{figure2b.ps}
\caption{The 180-Hz QPO (left) and the 280-Hz QPO (right) in XTE J1550--564. Upper panels: Power density spectra over the full PCA bandpass. Lower panels: Phase lag Fourier spectra for the $6.5-13.0$ keV and $< 6.5$ keV photons. The light-blue points show the frequency range over which we calculated the phase lags of the QPOs. Lines are the same as in Figure \ref{fig1}.}
\label{fig2}
\end{figure*}

\begin{figure*}
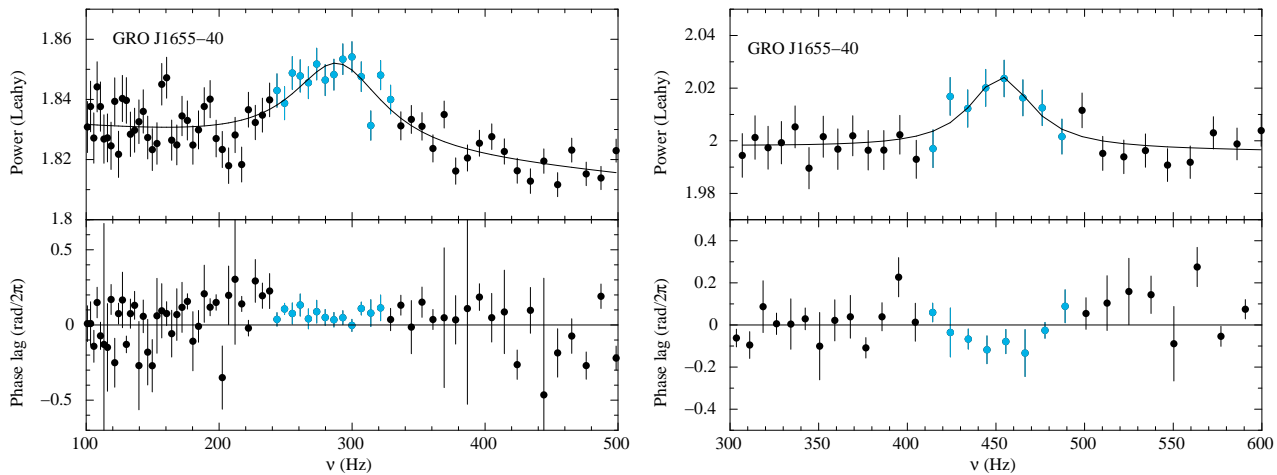

\centering
\includegraphics[width=0.35\textwidth,angle=-90]{figure3a.ps}
\includegraphics[width=0.35\textwidth,angle=-90]{figure3b.ps}
\caption{The 300-Hz QPO (left) and the 450-Hz QPO (right) in GRO J1655--40. Upper panels: Power density spectra over the full PCA bandpass (for the 450-Hz QPO using photons in the $13.0-26.3$ keV range). Lower panels: Phase lag Fourier spectra for the $13.0-15.2$ keV and the $< 13.0$ keV photons. The light-blue points show the frequency range over which we calculated the phase lags of the QPOs. Lines are the same as in Figure \ref{fig1}.}
\label{fig3}
\end{figure*}

In Table \ref{tab2} we give the average frequency (which we used as a reference to shift the power density spectra if we needed to apply the shift-and add technique) and Full-Width at Half-Maximum, $FWHM$, of the HFQPOs, and we show the best-fitting model to the power density spectra as a solid line in the upper panels of Figures \ref{fig1}--\ref{fig3}.

%
%
%

We then averaged the 16-s real and imaginary part of the Fourier transform as a function of Fourier frequency for each observation and each source separately. When necessary, as we did for the power density spectra, we used the average HFQPO frequency for each observation of a given source, we shifted the frequency scale of the Fourier transform of each full observation to bring the frequency of the HFQPO to that average value, and we then averaged the real and imaginary parts of the Fourier transforms \citep[see][]{avellar}

For each source (except IGR J17091--3624; see below) we selected the softest energy band in Table \ref{tab1} as the reference band, we calculated average cross spectra between each of the other energy bands in Table \ref{tab1} and this reference band and, following the description in \citet{vaughan-nowak} and \citet{nowak01}, we calculated the phase lags and the Fourier intrinsic coherence as a function of Fourier frequency for each of the cross spectra. In the case of GRS 1915+105 the softest band goes from channel 8 to channel 13, whereas for the other sources the softest band starts at channel 0 (see Table \ref{tab1}). To be able to compare the phase lags of the 67-Hz QPOs in, respectively, IGR J17091--3624 and GRS 1915+105, in the case of IGR J17091--3624 we also chose the band that goes from channel 8 to channel 13 as the reference band.\footnote{Since the phase lag is a relative quantity, the choice of the reference band only shifts the full phase-lag spectrum, but does not change the dependence of the lags with energy.} 

Finally, to calculate the phase lags of the HFQPO we averaged the phase lags over a frequency interval equal to the $FWHM$ of the corresponding HFQPO. Formally these are the phase lags of the light curves in the two energy bands used in the calculation, over the time-scale (the inverse of the Fourier frequency) of the HFQPO. For simplicity, in the rest of the paper we call these the phase lags of the HFQPO. 

%
%
%

%
%
%

Dead-time driven cross talk between different energy channels induces a $+\pi$ or $-\pi$ phase lag in the Poisson-noise dominated part of the power density spectrum. To correct for this, for each source we subtracted the cross spectrum in a frequency range well above that of the HFQPOs, in which the variability was dominated by Poisson noise. We confirmed that the cross spectrum in this frequency range has no significant imaginary part, as expected, and hence this procedure has no effect on the sign of the phase lags of the HFQPO \citep[see][for details]{vanderKlis:1987kh}. 

Since the broadband noise component in black-hole (and neutron-star) systems can extend up to a few hundred Hz \citep[e.g.,][]{sunyaev}, in the range where the HFQPOs in black-hole systems usually appear, we checked whether the phase lag of the QPO may be affected by the signal of this broadband component. For this, we instead subtracted the average cross spectrum of the broadband component over frequency intervals just below or above the QPO signal before we calculated the phase lags of the QPO. The measured phase lags did not change significantly when we did this compared to the case in which we subtracted the cross spectrum in the frequency range dominated by the Poisson noise. The error bars of the phase lags given in this paper account for the small differences between these procedures. 

\section{Results}
\label{res}

In Figure \ref{fig1} we show for GRS 1915+105 (left) and IGR J17091--3624 (right) the power density spectrum (upper panels) and the phase-lag Fourier spectrum (lower panels), concentrating on the frequency range that contains the HFQPOs. For the power density spectrum of GRS 1915+105 we used the full PCA band while for that of IGR J17091--3624 we took photons in the $2-25$ keV band. The phase lags were calculated for the $5.7-14.8$ keV and the $3.2-5.7$ keV photons. For GRS 1915+105 the green and light-blue points in both the upper and lower panels indicate the frequency range over which we measured the phase lags of the 35-Hz and 67-Hz HFQPOs, respectively. For IGR J17091--3624 the light-blue points in both the upper and lower panels indicate the frequency range over which we measured the phase lags of the 67-Hz HFQPO. From this Figure it is apparent that in GRS 1915+105, for this energy selection, the phase lags of the 35-Hz QPO are negative, whereas those of the 67-Hz QPO are positive, indicating soft and hard lags, respectively. For these two bands the phase lags of the 67-Hz QPO in IGR J17091--3624 are consistent with zero. This Figure also shows that the 67-Hz QPO in GRS 1915+105 is slightly asymmetric \citep[see also][]{belloni-altamirano2}. 

%
%
%
\begin{figure*}
\centering
\includegraphics[width=1\textwidth,angle=0]{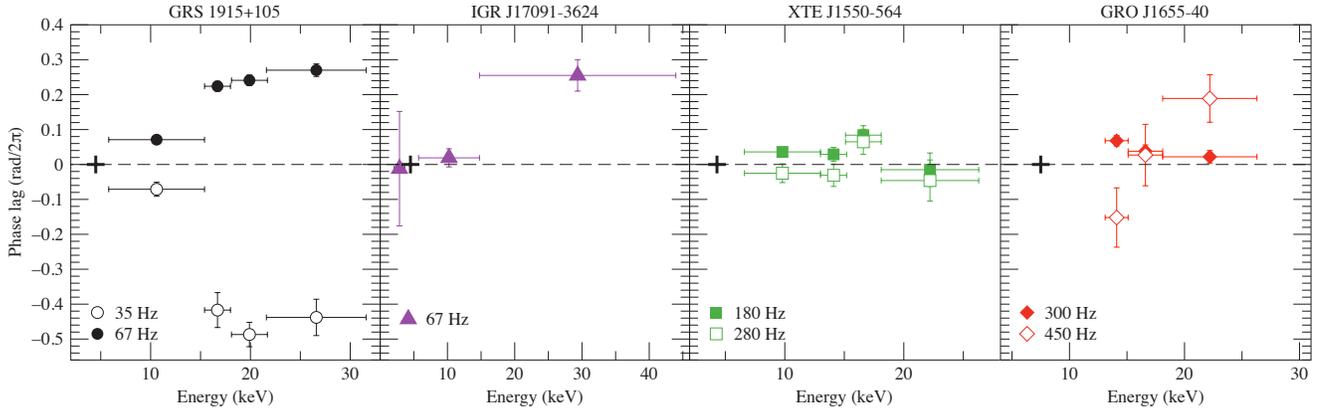}
\caption{Phase lags as a function of energy for the QPOs in GRS 1915+105 (left panel), IGR J17091--3624 (middle-left panel), XTE J1550--564 (middle-right panel) and GRO J1655--40 (right panel). The horizontal line across all panels shows the zero-lag level. The black cross shows the reference band.}
\label{fig4}
\end{figure*}

In Figure \ref{fig2} we plot the full-band power density spectrum (upper panels) and the phase-lag spectrum ($6.5-13.0$ keV relative to $< 6.5$ keV; lower panels) of, respectively, the 180-Hz (left) and 280-Hz (right) QPOs in XTE J1550--564. The light-blue points in all panels show the frequency range over which we measured the phase lags of these two QPOs. In XTE J1550--564, for this energy selection, the phase lags of the 180-Hz QPO are positive, whereas those of the 280-Hz are consistent with zero.

In Figure \ref{fig3} we plot, respectively, the full-band and the $13-26.3$ keV power density spectra (upper panels) and the phase-lag spectra ($13.0-15.2$ keV relative to $< 13.0$ keV; lower panels) of the 300-Hz (left) and 450-Hz (right) QPOs in GRO J1655--40. The light-blue points in all panels show the frequency range over which we measured the phase lags of these two QPOs. For this energy selection in GRO J1655--40 the phase lags of the 300-Hz QPO are positive, whereas those of the 450-Hz are (marginally) negative.

As described in \S\ref{obs}, we calculated the average phase lag over the frequency range covering one $FWHM$ of the QPO (Table \ref{tab2}), for all energy bands in Table \ref{tab1}. In the four panels of Figure \ref{fig4} we plot the phase lags of the QPOs of each of the four sources as a function of energy.

Figure \ref{fig4} shows that: (i) In GRS 1915+105 the phase lags of the 67-Hz QPO are hard and increase with energy, whereas the phase lags of the 35-Hz QPO are soft, and their magnitude increases with energy. From this Figure it is apparent that the 35-Hz and the 67-Hz QPOs have significantly different lag spectra. (ii) The lag spectrum of the 67-Hz QPO in IGR J17091--3624 is consistent with that of the 67-Hz QPO in GRS 1915+105. (iii) The lag spectra of the 180-Hz and the 280-Hz QPOs in XTE J1550--564 are independent of energy and are consistent with each other. (iv) In GRO J1655--40 the phase lags of the 300-Hz QPO are independent of energy, and the lag spectrum of this QPO is consistent with those of the 180-Hz and 280-Hz QPOs in XTE J1550--564. (v) The lag spectrum of the 450-Hz QPO is similar to those of the 67-Hz QPO in GRS 1915+105 and IGR J17091--3624. (vi) The lag spectra of the 300-Hz QPO in GRO J1655--40 and the 180-Hz and 280-Hz QPOs in XTE J1550--564 are significantly different from those of the 67-Hz QPO in GRS 1915+105 and IGR J17091--3624, the 35-Hz QPO in GRS 1915+105, and the 450-Hz QPO in GRO J1655--40. 

Since the frequency of the 300-Hz QPO in GRO J1655--40 changed by $\sim40$ Hz among the three observations that we combined to get the lags (see \S\ref{obs}), we also computed the phase lags of the HFQPOs for the three individual observations of this source, where the QPO frequency was, respectively, $\sim270$ Hz, $\sim290$ Hz, and $\sim310$ Hz. In all three cases the values and the energy spectrum of the phase lags are consistent, with large error bars, with each other and with those of the combined data, and inconsistent with those of the 450-Hz QPO in this source.

\begin{figure}
\centering
\includegraphics[width=0.35\textwidth,angle=-90]{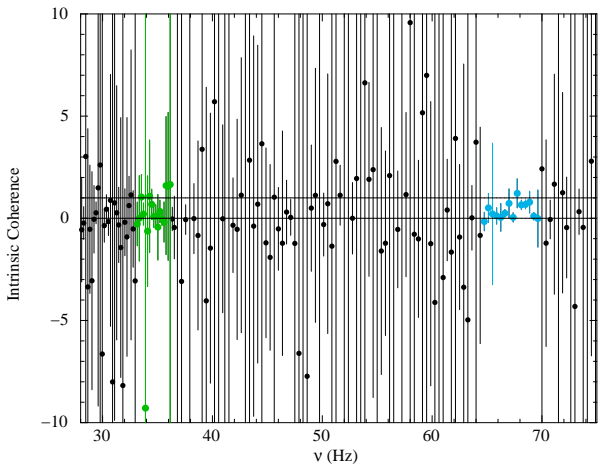}
\caption{Intrinsic coherence as a function of frequency for the 35-Hz and the 67-Hz QPO in GRS 1915+105. We calculated the intrinsic coherence for photons in the $5.7-14.8$ keV band relative to those in the $3.2-5.7$ keV band. The green and light-blue points indicate the frequency range of the 35-Hz and 67-Hz HFQPO, respectively. The horizontal lines indicate the level of coherence equal to zero and one, respectively.}
\label{fig5}
\end{figure}

We also calculated the intrinsic coherence of all the HFQPOs using pairs of light curves in two different energy bands. Except for a few combinations of energy bands, in all cases the intrinsic coherence of the HFQPOs across the QPO profile is rather noisy, and consistent both with 0 or 1 \citep[the possible extreme values of the intrinsic coherence; see][]{bendat01}. In Figure \ref{fig5} we show one of the examples in which the intrinsic coherence is well-defined across the QPO profile. In that Figure we plot the intrinsic coherence as a function of Fourier frequency for the 35-Hz and the 67-Hz QPOs in GRS 1915+105. To calculate the intrinsic coherence we used the photons in the $5.7-14.8$ keV and $3.2-5.7$ keV bands. The intrinsic coherence is well constrained (has small error bars) only around the frequency of the 35-Hz and the 67-Hz QPOs. Outside that range the errors increase dramatically. The other thing apparent from this Figure is that the average (across the QPO profile) intrinsic coherence is less than 1, although the individual measurements have in general large error bars, and are consistent with 1. Indeed, the error-weighted average of the intrinsic coherence of the 35-Hz and the 67-Hz QPOs using these two energy bands are $0.1\pm0.2$ and $0.3\pm0.1$, respectively. (We note, however, that if the individual measurements of the intrinsic coherence are not normally distributed, the weighted average could be biased.) A value of the intrinsic coherence lower than 1 implies that one or more of the following situations hold \citep[see][for details]{bendat01}: (i) The noise (due to the Poissonian nature of the signal) was not completely subtracted from either of the two signals, (ii) the light curves in the two energy bands are not linearly related or, if they are, (iii) the signal in one band is due to the signal in the other band plus at least one other uncorrelated signal in the system \citep[see][for details]{vaughan02}. We will not discuss this result further, and defer any possible interpretation to a future paper.

%
%
%

\section{Discussion}
\label{dis}

%
%

We present the first systematic study of the energy-dependent phase lags and coherence of the HFQPOs in the galactic black-hole systems GRS 1915+105, GRO J1655--40, XTE J1550--564, and IGR J17091--3624. Specifically, we studied the phase-lag spectrum of the 35-Hz and the 67-Hz QPOs in GRS 1915+1054, the 300-Hz and the 450-Hz QPOs in GRO J1655--40, the 180-Hz and the 280-Hz QPOs in XTE J1550--564, and the 67-Hz QPO in IGR J17091--3624. 

For the 67-Hz QPO in GRS 1915+105, the 450-Hz QPO in GRO J1655--40, and the 67-Hz QPO in IGR J17091--3624 the hard photons lag the soft ones, with the highest energy photons lagging the softest ones by up to $\sim0.5\pi$ rad. In the case of the 67-Hz QPO in GRS 1915+105, our results are consistent with those of \citet{cui}, based on another observation of this HFQPO. These are the first measurements of the phase lags of the 35-Hz QPO in GRS 1915+105 and the HFQPOs in the other three sources. The phase lags of the 35-Hz in GRS 1915+105 are negative, indicating soft lags, with the magnitude of the lags increasing with energy from $\sim0.1\pi$ rad at $\sim10$ keV up to $\sim\pi$ rad at $\sim25$ keV. For the 300-Hz QPO in GRO J1655--40, and the 180-Hz and the 280-Hz QPO in XTE J1550--564, the phase lags are independent of energy, and consistent with being zero or slightly positive (hard lags). In GRS 1915+105, the phase-lag spectrum of the 35-Hz QPO is inconsistent with that of the 67-Hz QPO. Similarly, the phase-lag spectrum of the 300-Hz and the 450-Hz QPOs in GRO J1655--40 are significantly different. On the contrary, the phase-lag spectrum of the 180-Hz and the 280-Hz QPOs in XTE J1550--564 are consistent with each other, and both are consistent with the phase-lag spectrum of the 300-Hz QPO in GRO J1655--40.

%
%

%
%

\cite{miller2001} reported the detection of two simultaneous HFQPOs in XTE J1550--564 at $\sim190$ Hz and $\sim270$ Hz. They detected the HFQPO at $\sim270$ Hz in each of six separate observations during the 2000 outburst of XTE J1550--564, whereas the HFQPO at $\sim190$ Hz is only detected, albeit at a single-trial significant of $3.5\sigma$, when they combine the six observations with the QPO at $\sim270$ Hz. Considering the number of trials reported by \citet{miller2001}, the QPO at $\sim190$ Hz is only $\sim2.3\sigma$ significant. Furthermore, from the analysis of all {\em RXTE} observations of XTE J1550--564 separately, \cite*{belloni-sanna-mendez} recently showed that there is only one significant HFQPO at a time, either at $\sim180$ Hz or $\sim280$ Hz. The similarity of their phase-lag spectrum suggests that the 180-Hz and the 280-Hz QPO in XTE J1550--564 are in fact the same QPO seen at different frequencies in different observations. This idea is reinforced by the fact that in GRS 1915+105 and GRO J1655--40, which show two well-identified simultaneous HFQPOs, the phase-lag spectrum of the HFQPO at the highest frequency is significantly different from that of the one at the lowest frequency (\S\ref{res}, Fig. \ref{fig4}). This comparison across sources is valid if the phase-lag spectrum of the highest (lowest) frequency HFQPO is independent of the QPO frequency or other source parameters, like black-hole mass or spin. If this interpretation is correct, XTE J1550--564 displays a single HFQPO that changes frequency by $\sim100$ Hz over different observations (it would not be the only case, though, since the 300-Hz QPO in GRO J1655--40 changes by $\sim40$ Hz as well), making HFQPO in black-hole systems more similar to kilohertz QPOs in neutron-star systems (see below). This also suggests that the lags could provide an alternative way to identify the high-frequency QPO features in galactic black-hole systems.

%
%

Given the frequency of the HFQPOs, the observed phase lags imply maximum time lags (in absolute value) ranging from $\sim14$ ms for the 35-Hz QPO in GRS 1915+105, to $\sim0.3$ ms for the 280-Hz QPO in XTE J1550--564 and the 300-Hz and the 450-Hz QPO in GRO J1655--40, with the magnitude of the maximum time lag decreasing more or less steadily with the frequency of the QPO. The small time lags imply a light-travel-time distance of $\sim100-4000$ km, or about $10-40$ gravitational radii for a 10-$\Msun$ non-rotating black hole. 

%
%

\subsection{Comparison with phase lags in active galactic nuclei}

\citet{zoghbi01} proposed that the $\sim30$-s soft time lags seen in the active galactic nucleus 1H0707--495 at a Fourier frequency of $\sim5\times10^{-4}$ Hz are due to reflection of photons from the corona around the central black hole off the accretion disc. Given the steep emissivity index inferred from the spectral fits to the time averaged spectra of this \citep{fabian} and other objects \citep[e.g.,][]{miller2009,chiang,risaliti}, the bulk of the reflection must take place very close to the inner edge of the accretion disc. 

The observed hard lags of the 67-Hz QPO in, respectively, GRS 1915+105 and IGR J17091--3624, and the 450-Hz QPO in GRO J1655--40 cast doubt on this interpretation, assuming that the same scenario holds for galactic black-hole systems and, that in each source, one of the HFQPOs reflects the Keplerian frequency at the inner edge of the accretion disc \citep[e.g.,][]{stella}. The only HFQPO in our sample that shows soft lags is the 35-Hz QPO in GRS 1915+105. If the frequency of this QPO is the Keplerian frequency at the inner edge of the accretion disc \citep[notice, however, that the lowest possible Keplerian frequency at the innermost stable circular orbit of a $10-\Msun$ black hole, \citealt{steeghs}, is $\sim200$ Hz; see also][]{belloni-altamirano2}, one would need to explain the existence of a simultaneous HFQPO still at higher frequencies, namely the 67-Hz QPO in this source. On the other hand, neither the 300-Hz QPO in GRO J1655--40, nor the 180-Hz/280-Hz QPO in XTE J1550--564 show the expected negative lags, making the scenario less likely. 

In the case of 1H0707--495 the lags become negative at $\sim5\times 10^{-4}$ Hz; if the frequency at which the lags become negative is proportional to mass (as in the case of lags produced at the inner edge of a Keplerian disc), this frequency would be $\sim250$ Hz for a 10-$\Msun$ black hole. This value is consistent with the frequency range of the HFQPOs in GRO J1655--40 and XTE J1550--56, but the time lags have the opposite sign. Furthermore, the magnitude of the time lags of the HFQPOs in these four systems is one to two orders of magnitude larger than the lags of 1H0707--495 scaled down by the black-hole mass-ratio. The $\sim14$ ms time lag of the 35-Hz QPO in GRS 1915+105 is consistent with the expected range, $5-150$ ms, from the extrapolation of the relation between soft time lags and black hole mass of \cite{demarco} for a $10$-$\Msun$ black hole. However, the extrapolation of the relation between frequency at which the soft time lags are observed and the black-hole mass of \citet{demarco} yields frequencies below $\sim1$ Hz, inconsistent with those of the HFQPOs in black holes.

\cite{middleton01} suggested that the QPO at $2.7 \times 10^{-4}$ Hz \citep{gierlinski} in the narrow-line type 1 Seyfert galaxy RE J1034+396 \citep[see][for a discussion on the significance of the QPO]{vaughan} could be the counterpart, scaled down in frequency, of the 67-Hz QPO in GRS 1915+105. However, the phase lags of the QPO in RE J1034+396 are soft and become softer as the energy increases \citep{middleton02}, whereas the lags of the 67-Hz QPO in GRS 1915+105 are hard and become harder as energy increases. Our results rule out that identification, unless the lag spectrum can switch from hard to soft  as the mass of the central black hole decreases by a factor of $\sim10^5$. In fact, the lag spectrum of the QPO in RE J1034+396 is very similar to that of the 35-Hz QPO in GRS 1915+105 (compare the right panel of Figure 9 in \citealt*{middleton02} and the left panel of Figure \ref{fig4}). If these two QPOs were the same feature scaled by mass, by analogy we would expect to see a QPO in RE J1034+396 at about twice the frequency of the $2.7 \times 10^{-4}$ Hz QPO, contrary to what is observed. 

%
%

\subsection{Phase lags and rms spectrum of the high-frequency QPOs}

The idea that the phase lags of the HFQPOs are due to reflection off the accretion disc also needs to address the rms spectrum of the HFQPOs. \citet*{Morgan:1997hn} and \cite{belloni-altamirano2} found that the rms spectrum (rms amplitude vs. energy) of the 67-Hz QPO in GRS 1915+105 increases very steeply with energy, from $\sim1$\% at 4 keV to $\simmore10$\% at 30 keV. Similarly, the rms spectrum of the 280-Hz QPO in XTE J1550--564 increases with energy \citep{miller2001}, and the 450-Hz QPO in GRO J1655--40 is only detected at high energies \citep{strohmayer}. In some cases, when HFQPOs are observed, the $2-25$ keV emission of the accretion disc is $\simless 10$ per cent of the total luminosity of the source \citep[e.g.,][]{remillard}, and therefore emission from the disc alone cannot explain the amplitude of the HFQPOs. If the oscillation mechanism that produces the HFQPOs takes place in the disc, the signal must be modulated and amplified in the corona \citep[see][for a similar discussion in the case of the kilohertz QPOs in neutron-star systems]{mendez04, sanna01}, and this should affect the phase lags of the HFQPO.

%
%

\subsection{Comparison with the broadband noise phase lags }

On time-scales of $10^{-2}-10^2$ s in black-hole (and neutron-star) X-ray binaries in the hard state the hard X-ray photons lag the soft ones by $\sim0.1$ rad, with the phase lags being more or less constant as a function of Fourier frequency \citep[e.g.][]{miyamoto01, nowak01, ford01}. These broadband frequency-independent hard phase lags are usually explained in terms of Compton up-scattering in a hot electron plasma close to the black hole: Low-energy photons gain energy in each scattering event; more scatterings imply higher energy of the emerging photons, and also a longer path length, therefore the higher-energy photons emerge later than the lower energy ones produced simultaneously. The phase lags of the 67-Hz QPO in, respectively, GRS 1915+105 and IGR J17091--3624, the 300-Hz and 450-Hz QPO in GRO J1655--40 and the 180-Hz/280-Hz QPO in XTE J1550--564 are consistent with this scenario, whereas the soft phase lags of the 35-Hz QPO in GRS 1915+105 contradict this idea. Soft phase lags could be due to hard photons emitted from the hot corona around the black hole that are Compton down-scattered by cold plasma in the accretion disc \citep[e.g.][]{reig, nobili,falanga01}. A Compton down-scattering mechanism for the time lags of the 35-Hz QPO in GRS 1915+105 is difficult to reconcile with the fact that the hard (power-law like)  component in the time-averaged spectrum of this system (and other black-holes systems) is interpreted as Compton up-scattering \citep{sunyaev01, payne01}. As shown by \citet{lin}, lags of opposite sign are possible even if the photon arrival times at different QPO harmonic frequencies are the same. This could perhaps explain the opposite sign of the possibly harmonically-related \citep{belloni-altamirano} QPOs at 35 Hz and 67 Hz in GRS 1915+105, without the need to resort to complex Compton up- and down-scattering mechanisms operating at the same time. 

%
%

\subsection{Comparison with the phase lags of high-frequency QPOs in neutron-star systems}

Neutron-star X-ray binaries also exhibit a pair of high-frequency QPOs, at frequencies between $\sim400-1200$ Hz, much higher than the frequency of the HFQPOs in black-hole systems. These are called kilohertz quasi-periodic oscillations (kHz QPOs). It is unclear, however, whether the kHz QPOs and the HFQPOs are the same phenomena, and whether the two members of the pair of kHz QPOs and those of the pair of HFQPOs (when two are present) can be identified with one another. From the pair of kHz QPOs, the one at lowest frequency, the lower kHz QPO, displays soft time lags of $\sim20-100\mu$s, with the magnitude of the lags increasing with energy \citep{vaughan01,vaughan02,kaaret01,avellar,barret13}. The time lags of the other kHz QPO, the upper kHz QPO, are hard, of $\sim10\mu$s, independent of energy, and significantly different from the time lags of the lower kHz QPO \citep{avellar}. 

The lags in neutron-star systems have similar properties to those in black-hole systems. For instance, in GRS 1915+105 the phase lags of the 35-Hz QPO and those of the lower kHz QPO in the neutron-star systems 4U 1608--52 and 4U 1636--53 are soft, whereas the phase lags of the 67-Hz QPO and those of the upper kHz QPO in 4U 1608--52 and 4U 1636--53 are hard. Furthermore, in GRS 1915+105 the absolute value of the time lags of the 35-Hz is about 3 times that of the 67-Hz QPO ($\sim14$ ms and $\sim5$ ms, respectively), while in 4U 1636--53 the absolute value of the time lags of the lower kHz QPO is about twice that of the upper kHz QPO ($\sim20\mu$s and $\sim10\mu$s, respectively). Based on this, it is tempting to identify the lower and upper kHz QPO in neutron stars with the 35-Hz and the 67-Hz QPO in GRS 1915+105, but this identification is not without problem: The phase lags of the 67-Hz QPO in GRS 1915+105 increase with energy, whereas in the two neutron-star systems the time lags of the upper kHz QPO are independent of energy, similar to what we observe for the phase lags of the 180-Hz and 280-Hz QPO in XTE J1550--564 and the 300-Hz QPOs in GRO J1655--40. Additionally, GRS 1915+105 shows QPOs at 27 Hz \citep*{belloni-mendez-sanchez} and 41 Hz \citep{strohmayer02} in observations in which the 67-Hz QPO is also present, which makes the identification with the pair of kHz QPOs in neutron stars more difficult. 

Notice, by the way, that if the same mechanism proposed to explain the lags in 1H0707--495 is responsible for the lags in galactic black holes and neutron-star systems, the magnitude of the time lags of the kHz QPOs and those of GRS 1915+105 imply that the distance between the source of photons and the reflector in GRS 1915+105 is $500-700$ times larger than in the two neutron-star systems 4U 1608--52 and 4U 1636--53, whereas the mass ratio is only $\sim5$.

%
%

\citet{lee02} proposed a Compton up-scattering scenario that can explain the soft time lags of the lower kHz QPO in neutron-star systems. They found a solution of the linearised Kompaneets equation \citep{kompaneets} in which the disc and the corona are coupled and exchange energy, and the coronal and disc temperatures oscillate coherently at the QPO frequency \citep[see also][]{lee01}. This model reproduces both the time lags \citep{vaughan01,vaughan02} and rms spectrum \citep{berger01} of the lower kilohertz QPO in 4U 1608--52. Also, since the resonance is restricted to a limited frequency range, the idea of \cite*{lee02} is consistent with the hard lags on time-scales of $10^{-2} - 10^{2}$ s that we previously mentioned and, in the case of neutron stars, the hard lags of the upper kilohertz QPO \citep{avellar}. This mechanism may also apply to accreting black-hole systems; it remains to be seen whether this could explain as well the soft lags of the 35-Hz QPO and the hard lags of the 67-Hz QPO in GRS 1915+105.

\section*{Acknowledgments}

The authors wish to thank Guobao Zhang for useful discussions and Matt Middleton for comments on the first version of the manuscript. We are grateful to an anonymous referee for suggestions that helped us clarify some aspects of this paper. This research has made use of data obtained from the High Energy Astrophysics Science Archive Research Center (HEASARC), provided by NASA's Goddard Space Flight Center. This research made use of NASA's Astrophysics Data System. MM wishes to thank MNM.

\label{lastpage}

\end{document}